# The Role of Shift Vector in High-Harmonic Generation from Non-Centrosymmetric Topological Insulators under Strong Laser Fields

Chen Qian,[1,2] Chao Yu,[1] Shicheng Jiang,[5] Tan Zhang,[2,3] Jiacheng Gao,[2,3] Shang Shi,[2,3] Hanqi Pi,[2,3] Hongming Weng,[2,3,4] and Ruifeng Lu[1,2,*]

[1] *Institute of Ultrafast Optical Physics, Department of Applied Physics, Nanjing University of Science and Technology, Nanjing 210094, P R China*
[2] *Institute of Physics, Chinese Academy of Sciences, Beijing National Laboratory for Condensed Matter Physics, Beijing 100190, P R China*
[3] *School of Physics, University of Chinese Academy of Sciences, Beijing 100049, China*
[4] *Songshan Lake Materials Laboratory, Dongguan, Guangdong 523808, China*
[5] *State Key Laboratory of Precision Spectroscopy, East China Normal University, Shanghai 200062, China*

## Abstract

As a promising avenue to obtain new extreme ultraviolet light source and detect electronic properties, high-harmonic generation (HHG) has been actively developed in both theory and experiment. In solids lacking inversion symmetry, when electrons undergo a nonadiabatic transition, a directional charge shift occurs and is characterized by shift vector, which measures the real-space shift of the photoexcited electron and hole. For the first time, we have revealed that shift vector plays prominent roles in the real-space tunneling mechanism of three-step model for electrons under strong laser fields. Since shift vector is determined by the topological properties of related wave functions, we expect HHG with its contribution can provide direct knowledge on the band topology in noncentrosymmetric topological insulators (TIs). In both Kane-Mele model and realistic material BiTeI, we have found that the shift vector reverses when band inversion happens during the topological phase transition between normal and topological insulators. Under oscillating strong laser fields, the reversal of shift vector leads to completely opposite radiation time of high-order harmonics. This makes HHG a feasible all-optical strong-field method to directly identify the band inversion in non-centrosymmetric TIs.

---

[*] rflu@njust.edu.cn

# I. INTRODUCTION

The ultra-fast and intense laser allows manipulation and measurement of atomic-size electron motion on an attosecond time scale [1]. A major branch of attosecond and strong-field physics is high-harmonic generation (HHG), which can be utilized to achieve isolated attosecond pulses and detect dynamics of electrons [2]. Since the HHG was observed by Ghimire *et al*. in ZnO crystal [3], a series of experimental findings of solid HHG have subsequently emerged and correspondingly the microscopic theory has been improving gradually [4-7]. Due to the periodic lattice arrangement, dense electron concentration and macroscopic symmetry of solid-state materials, many novel phenomena different from HHG in gas phase have been observed in solids, such as the band-gap dependent cut-off energy [8], dynamic Bloch oscillation [9,10], anomalous velocity induced by Berry curvature [11-14], and orientation dependence of harmonics determined by crystalline symmetry [12,15-17]. So far, nevertheless, the theory of solid HHG has not yet been fully developed to a thorough stage [18]. There are still many unexplored or unsolved problems that are of great and broad interest.

It is universal that the electron transits among energy levels under the light irradiation with appropriate frequencies. In the past few years, intensive attentions have been paid to the role of the energy eigenvalue, whilst little to the possible displacement of the electron wave packet during the transition driven by lights [8,19,20]. Most approaches in theory are reasonable for gaseous medium or crystals with inversion, because in either case there is no directional flow of electrons under photoexcitation [21]. However, in a solid material lacking inversion symmetry [22], the electron wave packet tends to shift in a certain direction during interband transition, and the involved physical process is described by a shift vector, which measures the displacement between charge centers of different bands. Previously, the knowledge is limited to the shift current based entirely on the perturbation theory. It is a second-order nonlinear process used to explain the microscopic mechanism of the bulk photovoltaic effect [23-29]. Due to the existence of shift vector in non-centrosymmetric system, nonperturbative high-harmonic pulse trains (HHPTs) in strong laser fields exhibit a property of asymmetry in the time domain: the burst period of the HHPTs is modified from half an optical cycle to the whole single cycle. Herein, we apply this discovery to the non-centrosymmetric topological insulators (TIs), a type of quantum material with large shift vectors [26,29].

As a new class of quantum matter with time reversal symmetry, TIs have exotic characteristics with insulating bulk band and metallic surface state. Their robust surface state is protected by topological invariants of the bulk state and supports quantized spin current against backscattering [30-34]. Therefore, TIs are expected to be applied to dissipationless devices, quantum spintronics and quantum computing [31,33]. It is inspiring that high-order nonlinear optical studies on TIs are continuously progressing, especially for the explorations on topological phase discrimination and unique electronic dynamics [35-40]. The topological phases are distinguished experimentally mainly by the topological surface state appeared in the atomic layers on the surface [41-43], whereas the formation of the topological surface state is completely determined by the bulk state. Therefore, it is necessary to trace the root. Substantially, the topological phase transition features band inversion of the bulk state [31,32,34,44,45], but which is quite difficult to be directly detected in experiments. An essential reason is that the band inversion is mainly indicated by the exchange of the components of wave functions, and an observable physical quantity that can identify it is lacking. Nonlinear optical techniques, especially HHG, have advantages in directly detecting the relationship of two or more wave functions and being strictly constrained by their symmetries [12,15-17]. We claim that band inversion can be recorded by temporal high-harmonic spectrum instead of conventional techniques such as angle-resolved photoemission spectroscopy [42,43]. Thus, we propose an experimentally feasible method of using ultrafast and intense lasers to characterize the band inversion in non-centrosymmetric TIs, which is accompanied by the reversal of the shift vector that can modify the HHPTs.

In this paper, we first use the Kane-Mele (KM) model [30], which is a paragon of the quantum spin Hall effect, to perform semi-classical dynamics analysis and generalized numerical calculations. We then extend the model to a layered polar material BiTeI [26,45] to further confirm the practical significances of our conception. Both the KM model and BiTeI crystal can be controlled to experience the topological phase transition by lattice modulation. By addressing the role of the shift vector in HHG from non-centrosymmetric TIs, we are aiming not only to give an improved physical picture of the semi-classical three-step model for solid HHG, but also to provide an all-optical strong-field method for detecting the topological phase transition in experiments.

## II. THEORY

In this work, the theoretical proposals and numerical simulations are based on the semiconductor Bloch equations (SBEs) [18,46-48] in Houston representation (atomic units are used throughout unless otherwise stated),

$$i\partial_t \rho_{nm}^{k(t)}(t) = \left[\varepsilon_m^{k(t)} - \varepsilon_n^{k(t)} - \frac{i(1-\delta_{nm})}{T_2}\right]\rho_{nm}^{k(t)}(t) - E(t) \cdot \sum_l \left[d_{ln}^{k(t)}\rho_{lm}^{k(t)}(t) - d_{ml}^{k(t)}\rho_{nl}^{k(t)}(t)\right], (1)$$

which indicates the evolution of electron density $\rho_{nm}^k(t)$ with time $t$ under laser fields. The time dependent crystal momentum $k(t) = K + A(t)$ shows the adiabatic evolution path of electrons in the Brillouin zone (BZ), where $K$ represents quasi-crystal canonical momentum and $A(t)$ is the vector potential of the laser field $E(t)$. $\varepsilon_n^k$ is the energy of the $n$th band at $k$ point and the dephasing time $T_2$ is determined by the relaxation time of the system. The dipole matrix element $d_{nm}^k = i\langle u_{n,k}|\nabla_k|u_{m,k}\rangle$ is composed of the periodic part of Bloch wave function $|u_{n,k}\rangle$, and can be the adiabatic Berry connection ($n = m$) or diabatic transition dipole matrix element ($n \neq m$).

To describe the role of the shift vector in HHG, we start from interband harmonics under saddle point approximation which follows from Eq. (1) [46,47,49,50]. More details on how to calculate the current are shown in Appendix A. In the case of two-band approximation and considering that only a few of electrons can transit from the valence band to the conduction band, the interband current parallel to the laser polarization direction in frequency domain can be written as (ignoring the dephasing time, and see Appendix B for the derivation)

$$J_{inter}(\omega) = \sum_{K \in BZ} \int_{-\infty}^{+\infty} dt \int_{-\infty}^{t} dt'\, \varepsilon_g^{k(t)} \left|d_{vc}^{k(t)}\right| E(t') \left|d_{cv}^{k(t')}\right| e^{-i[S(K,t,t') - \omega t]} + c.c., \quad (2)$$

where $\varepsilon_g^k = \varepsilon_c^k - \varepsilon_v^k$ is the energy difference between the conduction band and valence band, and the accumulated phase from $t'$ to $t$ is

$$S(K, t, t') = \int_{t'}^{t} \varepsilon_g^{k(\tau)}\, d\tau - \int_{k(t')}^{k(t)} \left[d_{cc}^{k(\tau)} - d_{vv}^{k(\tau)}\right] dk(\tau) - \left[\phi_{vc}^{k(t)} + \phi_{cv}^{k(t')}\right], \quad (3)$$

with $\phi_{nm}^k$ being the transition dipole phase (TDP) [51]. Taking the saddle point analysis for $S(K, t, t') - \omega t$,

$$\partial_{t'} S(K, t, t') = -\varepsilon_g^{k(t')} - E(t') \cdot R_{cv}^{k(t')} = 0, \quad (4a)$$

$$\nabla_k S(K, t, t') = \int_{t'}^{t} \nabla_k \varepsilon_g^{k(\tau)}\, d\tau - R_{cv}^{k(t)} + R_{cv}^{k(t')} = 0, \quad (4b)$$

$$\partial_t S(K, t, t') = \varepsilon_g^{k(t)} + E(t) \cdot R_{cv}^{k(t)} = \omega, \quad (4c)$$

with $R_{cv}^k = d_{cc}^k - d_{vv}^k - \nabla_k \phi_{cv}^k$ being the shift vector, which is a combination of the Berry connections and TDP of the conduction band and valence band. The gradient of the TDP cancels out the gauge dependence of the Berry connections, so the shift vector is gauge invariant. Eqs. (4a)-(4c) are the formulations of the semi-classical three-step model. As far as we know, the influence of the shift vector $R_{cv}^k$ that exists in these three formulas has been overlooked so far [46,50,51].

The Berry connection forms the shift vector and here can be understood as the real-space charge center with quasi-momentum $k$ in a certain band. For the system with broken inversion symmetry, the charge centers of different bands usually do not coincide (see Appendix C for more discussions). The electron wave packet subject to laser field undergoes interband transition with quasi-momentum $k$, which appears as a hopping from one charge center to another in real space. This instantaneous process is described by the shift vector. Usually, this physical quantity is given by the second-order nonlinear process under the perturbation theory [23-25]. Obviously, in a strong field, no matter how excitation and recollision happen, the process that originates from the shift vector should also exist. We can see this process by solving Eqs. (4a)-(4c), which denotes a new real-space three-step model as shown in Fig. 1: (1) At time $t'$, the electron localized in the valence band charge center (VBCC) is excited by the laser to the conduction band charge center (CBCC) with the displacement $R_{cv}^{k(t')}$, tunneling through the combined barrier of Coulomb potential and laser electric field, and leaving a hole in the valence band. (2) The excited electron-hole pair is accelerated along the valence and conduction bands, respectively, by the oscillating laser field, and the HHG is produced by rapidly changing currents. (3) At time $t$, the electron is pulled back from the CBCC to the VBCC by Coulomb potential and electric field with the displacement $R_{cv}^{k(t)}$, emitting photons of energy $\omega$ with the modification from the displacement.

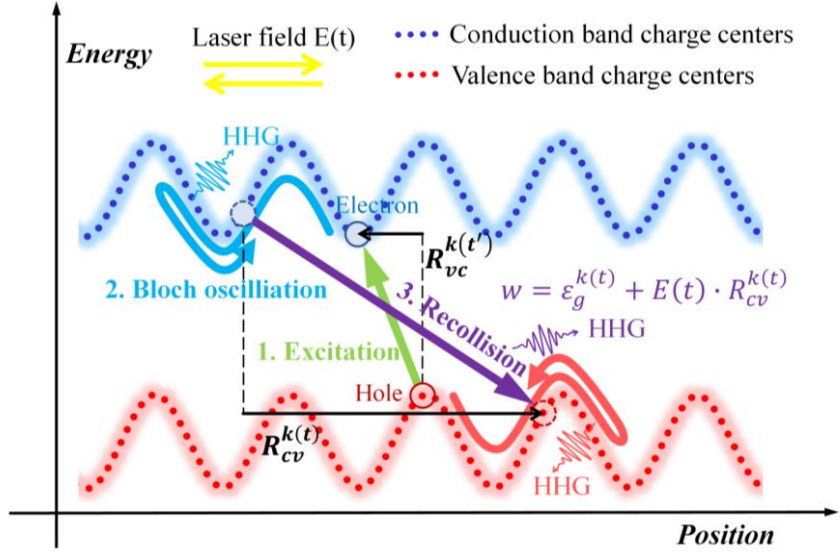

FIG. 1. Three-step model in real space for non-centrosymmetric systems under strong laser fields $E(t)$ with contributions from the shift vector $R_{cv}^k$ to HHG. The electron and hole pairs undergo three steps including tunneling excitation (green), Bloch oscillation (blue and red) and recollision (purple), transferring among different charge centers marked with blue and red dots in the wave-like dotted lines. In addition to variation in energy, the positions of the electron-hole pair in real space change with the oscillating laser field $E(t)$. The displacements of the electron at the time of excitation and recollision can be described by the shift vectors, as the black arrows indicate.

Since the single-electron effect is considered, according to Eq. 4(b), the displacements of the electron and hole are exactly coincident at the recollision moment, which seems to suggest that they should create and annihilate in pairs. Nevertheless, we only pay attention to the directional shift when the electron tunnels, and it doesn't matter here whether the electron recollides back with the previous hole in a precise way or not [3,50,52].

## III. THE CASE BASED ON KANE-MELE MODEL

### A. Equilibrium state analysis

We further confirm the mentioned process using the KM model that hosts large shift vectors. This model is characterized by a $Z_2$ topological index that distinguishes the quantum spin Hall insulator phase from the normal insulator phase with time reversal symmetry [30]. The KM model is generalized from the tight-binding Hamiltonian of graphene with the nearest hopping term $t$, with the spin-orbit coupling (SOC) term $\lambda_{SO}$ opening a gap at the Dirac points $K$ and $K'$, and a staggered sublattice potential term $\lambda_v$ breaking the inversion symmetry. The simultaneous existence of the SOC and broken inversion symmetry results in a spin splitting of bands at generic $k$-points. For simplicity, we do not consider the Rashba term $\lambda_R$, which can arise from an external electric field in perpendicular direction or interaction with a substrate. For such case, mirror symmetry about the $xy$ plane remains and the spin in the perpendicular direction $S_z$ is a good quantum number, so we can decompose the KM model into two $2 \times 2$ Hamiltonians with opposite spin,

$$H_{\uparrow(\downarrow)}(k) = d_1(k)\sigma_x + d_{12}(k)\sigma_y + d_2\sigma_z \pm d_{15}(k)\sigma_z, \tag{5}$$

where $\sigma_{x,y,z}$ is the pauli matrices, the arrows denote the spin direction, the substitutable plus and minus are specified for the spin-up and spin-down cases, respectively, which are independent of each other.

$$d_1(k) = t(1 + 2cos(x)cos(y)), \tag{6a}$$

$$d_{12}(k) = 2t(cos(x)sin(y)), \tag{6b}$$

$$d_2 = \lambda_v, \tag{6c}$$

$$d_{15} = \lambda_{SO}(2sin(2x) - 4sin(x)cos(y)), \tag{6d}$$

in which $x = \frac{1}{2}a_0 k_x$, $y = \frac{\sqrt{3}}{2}a_0 k_y$, and $a_0$ is the lattice constant.

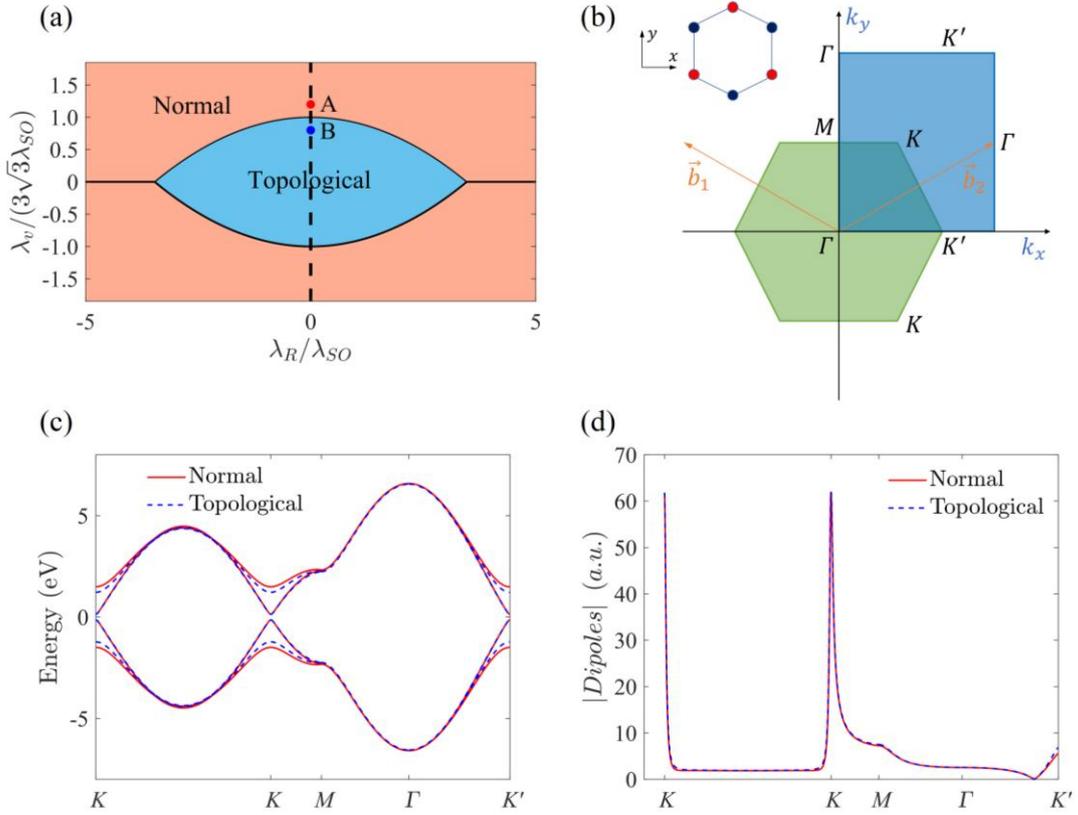

FIG. 2. Kane-Mele (KM) model in two phases. (a) The phase diagram of the KM model for $t = 0.08$, $\lambda_{SO} = 0.06t$, lattice constant $a_0 = 5.16$, where A point (red, $\lambda_v/(3\sqrt{3}\lambda_{SO}) = 1.2$) is in the normal phase and B point (blue, $\lambda_v/(3\sqrt{3}\lambda_{SO}) = 0.8$) is in the topological phase. The two points are both on the dashed line with $\lambda_R = 0$. (b) The Brillouin zone (BZ), where the orange arrows denote reciprocal lattice vectors, the first BZ is depicted by the green regular hexagon, which can also be converted to the blue rectangle. The inset shows the real-space honeycomb lattice with two different kinds of atoms (red and blue) (c) The band structures and (d) transition dipole amplitudes (TDAs) at the A and B points, which have the same bandgap of 0.27 eV and maximal transition dipoles at $K$ point in the BZ, here we only plot the TDAs between the two spin-down bands that have minimal gap at the $K$ point.

As shown by the dashed line in Fig. 2(a), we take $\lambda_R = 0$, and modify $\lambda_v$ to control the topological phase transition. This system is in the normal phase (NP) when $|\lambda_v|/(3\sqrt{3}\lambda_{SO}) > 1$, and in the topological phase (TP) when $|\lambda_v|/(3\sqrt{3}\lambda_{SO}) < 1$ [Fig. 2(a)].

As the topological phase transition approaching, the bandgap gradually tends to close, and then the wave functions of valence and conduction bands have an exchange when the gap opens again. This is known as band inversion, which causes the shift vector composed of the wave functions to reverse in the band inversion region [Fig. 3]. We can also see this from the expression of the shift vector,

$$R_{cv}^k \xrightarrow{band\ inversion} R_{vc}^k = d_{vv}^k - d_{cc}^k - \nabla_k \phi_{vc}^k = d_{vv}^k - d_{cc}^k + \nabla_k \phi_{cv}^k = -R_{cv}^k, \qquad (7)$$

where the band inversion results in an exchange of band indices, and the transformation of TDP is due to the fact that the transition dipole is a Hermitian operator.

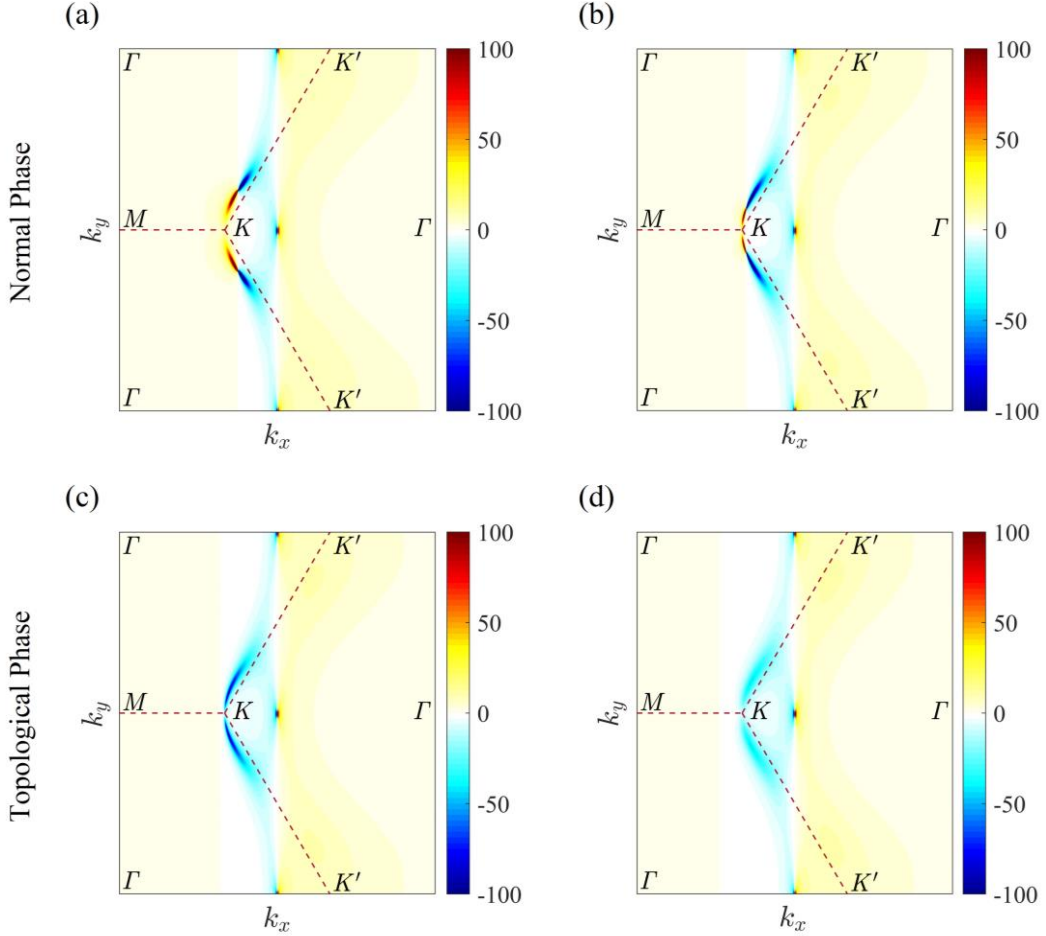

FIG. 3. Shift vectors along the $k_y$ direction (the applied laser field is in the same direction) in the first BZ (see the blue rectangle in Fig. 2(b)) of the KM model with (a) $\lambda_v/(3\sqrt{3}\lambda_{SO}) = 1.2$, (b) $\lambda_v/(3\sqrt{3}\lambda_{SO}) = 1.05$, (c) $\lambda_v/(3\sqrt{3}\lambda_{SO}) = 0.95$, and (d) $\lambda_v/(3\sqrt{3}\lambda_{SO}) = 0.8$. In the vicinity of the $K$ point, the shift vector peak is positive in the normal phase ((a), (b)) and negative in the topological phase ((c), (d)), but it is negative in all cases away from the $K$ point.

Since the KM model satisfies time reversal symmetry and the shift vector is invariable under the time reversal (see Appendix D for a derivation), we have $R_{cv,\uparrow}^k = R_{cv,\downarrow}^{-k}$ (the arrows denote the spin direction), which means only the two spin-down bands are needed to be considered, with the spin up ones being their time reversal pairs. Figure 3 shows that the peaks of the shift vector are localized near the $K$ point with the maximal transition dipole amplitudes (TDAs). In the trivial phase, shift vector peaks around the $K$ point have abrupt reversals [Figs. 3(a) and 3(b), see the color change from red to blue along $K-K'$ path], while the sign of the peaks is not changed in the topological phase [Figs.

3(c) and 3(d)]. The magnitude of peaks is much larger than the lattice constant of the system, and some prominent effects are expected.

**B. Analysis of KM system in laser fields**

We select two representative phase points A and B in the KM model to carry out numerical calculations over the entire BZ using Eq. (1). As marked in Fig. 2(a), A and B points are in the NP and TP, respectively (we specify them as NP and TP points later). These two phase points own the similar band structures (the same bandgap of 0.27eV) [Fig. 2(c)] and TDAs [Fig. 2(d)]. As shown in Figs. 3(a) and 3(d), the significant difference between the two phase points is the shift vectors, which are opposite to each other due to the band inversion near the bandgap, and where the transition probability is maximal. We choose an intense mid-infrared laser polarized linearly in the $k_y$ direction with the wavelength $\lambda = 6.7$ μm, the peak strength of electric field $E_0 = 7.18$ MV/cm, and the full-width at half-maximum of 160fs in a Gaussian envelope [Fig. 4(a)]. Under such laser parameters, nonadiabatic tunneling excitation dominates according to Keldysh parameter $\gamma \approx 1$ that the electronic state can barely change. The dephasing time we selected is a quarter of one laser cycle, and harmonics parallel to laser field are taken into consideration. Figure 4(b) plots the high-harmonic spectra of the NP and TP points. A clear platform structure and cutoff can be observed. Their frequency spectra show almost identical structures due to their similar band structures and TDAs [36], and the essential differences can hardly be revealed. The odd and even harmonics both appear in the spectra due to the inversion symmetry breaking. The low-order area is dominated by the odd-order harmonics, while in the high-order area the efficiencies of odd- and even-order harmonics are comparable. We notice that the higher-order harmonics deviate significantly from the integer orders, which results from their special band structures similar to Dirac cones and the model parameters we used here (see Appendix F for more discussions).

Since it is difficult to get prominent information only from frequency domain, we focus on the HHPTs for the NP and TP points [Figs. 4(e) and 4(f)], which come from the integration of time-frequency sprectrograms for each high-order harmonic [Fig. 4(c)]. Here, we show the HHPTs for the 20th and 25th orders in particular (see Appendix G for some other orders). Firstly, we can see that the HHPTs for both phase points burst once in each optical cycle around the peak of the electric field. The reversal of the laser field results in distinctly different radiation intensity of high-order harmonics in the adjacent half cycle, and the asymmetric HHG in time domain is obtained. Secondly, we find that the HHPTs emitting from the NP and TP points are locked to the positive and negative peaks of laser fields, respectively, so they have completely opposite radiation time.

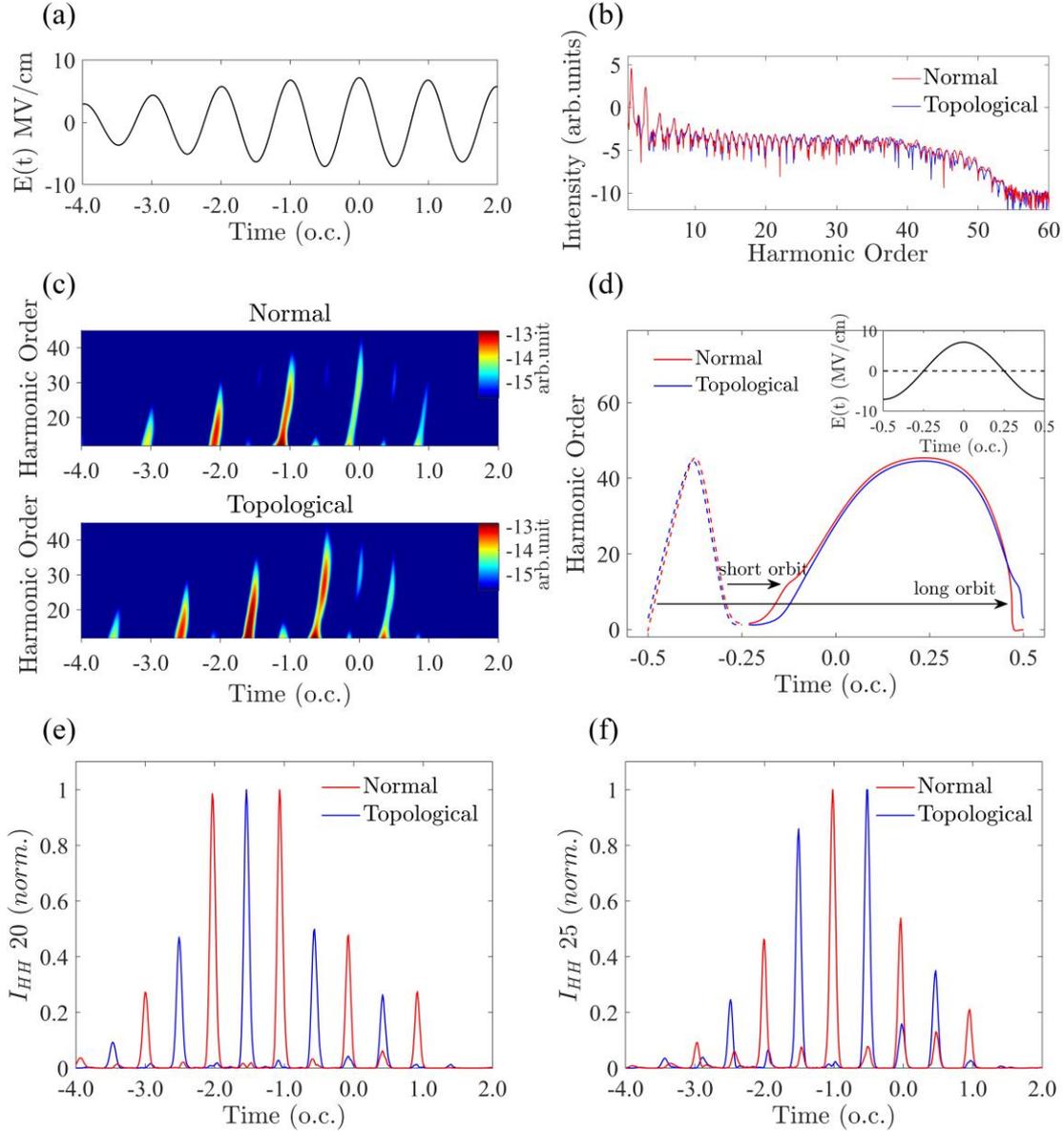

FIG. 4. HHG from normal (NP) and topological phases (TP). (a) The waveform of the mid-infrared driving field with six main cycles are shown. (b) High-harmonic spectra of the NP and TP points. (c) The time-frequency sprectrograms of the NP and TP points (the color scale is logarithmic). (d) The saddle point analysis as a function of birth time (dashed lines) and recollision time (solid lines) calculated by the saddle point equations, the inset shows the laser field strength in a single cycle. (e) The normalized intensity of the 20th and (f) 25th order high-harmonic pulse trains (HHPTs) extracted from (c) to show their temporal fine structures.

To explain the obtained interesting HHPTs, a qualitatively reasonable semi-classical analysis is critical. For the systems we calculate, the higher-order harmonics are dominated by the interband contribution [17,46] and burst periodically (see Appendix F for more details). Therefore, the saddle point analyses for interband currents based on Eqs. (4a)-(4c) is used here. Considering the time reversal symmetry of the KM model, we select two spin-down bands to proceed calculations for the NP and TP points, and identical results can be obtained in the spin-up case.

As we discussed before, the tunneling displacement determines the instantaneous change of electron potential in the laser field, and meanwhile, the recollision electron converts the potential change into the variation of emitted photon energy. The tunneling displacement is denoted by the shift vector.

From Eq. (4c), we can conclude that the energy of the emitted photon is determined not only by the difference between transition energy levels, but also by the coupling of the strong laser field with the shift vector. The results based on saddle point approximation in Fig. 4(d) show two distinct differences of about 10 harmonic orders. One is from the short orbit and the other is from the long orbit. Both of them originate from the reversal of shift vector. Since the electric field is positive when the short orbit recollides, the radiated photon energy of the NP point is higher than that of the TP point. The electric field is negative and the situation is inversed when the long orbit recollides. Therefore, the high-energy photons emitted from the NP and TP points are locked to the positive and negative electric fields, respectively.

More to the point, we demonstrate that the temporal asymmetry of HHPTs is induced by the reversal of electric fields, and it is necessary to ensure that the effect from the shift vector cannot be extinguished by the inversion symmetry or others (see Appendix C). The temporal asymmetry is determined by crystal symmetry breaking, which is independent of band inversion or topological properties. However, the completely opposite radiation time of HHPTs for the two phase points is owing to the reversal of the shift vector, which originates from the band inversion.

This temporal asymmetric phenomenon is similar to Refs. [16,53], in which the transformation of the HHG burst cycle is induced by the crystal symmetries in different directions. This suggests that the temporal profile of HHPTs can be resolved in experiments.

To quantitatively describe and distinguish the burst time of HHPTs, we introduce a normalized difference called tropism as

$$T_n = \frac{I_n^+ - I_n^-}{I_n^+ + I_n^-}, \tag{8}$$

where $n$ is the harmonic order, $I_n^+$ and $I_n^-$ denote the intensity of the emitted photon when electric field is positive and negative, respectively. The tropism is used to represent the dominate radiation time of HHG relative to the positive or negative laser field.

Due to the completely opposite radiation time of HHPTs between different phases, they should have opposite HHG tropisms. According to this, we plot the phase transition process of the KM model (without Rashba interaction) by an all-optical strong-field method [Fig. 5]. Notice that, there is an additional sign reversal before and after the inversion symmetry approaching, i.e., around $\lambda_v = 0$, because at this time the shift vector also reverses its direction due to the spatial inversion operation. As a result, the sign of the HHG tropism will reverse when the system undergoes the spatial inversion or the band inversion. When the SOC is doubled, we can also get a similar result in Fig. 5(b), which means that the tropism has little dependence on the strength of SOC. In the vicinity of the phase transition, we can also see that the tropisms vanish for higher order harmonics, which results from that the shift vector peak is confined to the region with a small bandgap [Fig. 3]. In fact, we can merely focus on the sign reversal of tropism to identify topological phase transitions. Thus, the experimental observation of the temporal profile of HHPTs is adequate for this scheme.

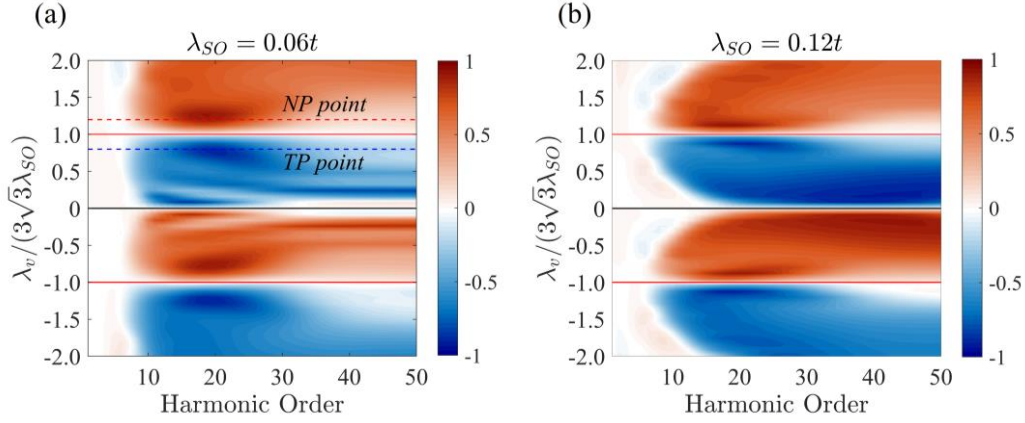

FIG. 5. HHG tropisms to the laser field in the KM model system, as a function of $\lambda_v/\lambda_{SO}$ and harmonic orders for (a) $\lambda_{SO} = 0.06t$, and (b) $\lambda_{SO} = 0.12t$. All the other parameters are the same. There are three sign changes in the diagrams. The middle one is from the spatial inversion operation (black solid line). The other two are from the topological phase transitions (red solid lines). The tropism spectra of the selected NP point (red dashed line) and TP point (blue dashed line) are indicated in (a).

In this paper, we do not pay attention to the harmonics perpendicular to the driving field caused by the Berry curvature. Although the fact that the Berry curvature also reverses the sign as the band inversion occurs [36], it is an odd function in the BZ of time reversal systems, and its reversal has no practical effect on HHG as the integral of the entire BZ.

## IV. THE CASE OF REAL MATERIAL: BiTeI

To illustrate the practical significance of shift vector and extend the conception to real materials with Rashba-like SOC interaction, we apply above methods and arguments to a layered polar material BiTeI [Fig. 6(a)]. The multiple layers are connected by the Van der Waals interaction. Under ambient pressure, BiTeI is a normal insulator [45]. It has a pressure-tunable bandgap located near $A$ ponit, the center of orthohexagonal boundary surface of the BZ [Fig. 6(b)]. The system is a Weyl semimetal when the bandgap closes during the topological phase tranistion [54], and the inversion symmetry is always broken. The bands of BiTeI have a Rashba-type splitting in the BZ except the time reversal invariant momenta [Figs. 6(c)-6(e)], which is aligned with the case in the KM model.

Our first-principles calculations were performed by Vienna ab initio simulation package (VASP) [55] based on density functional theory. The bulk energy bands and charge density of three-dimensional BiTeI were calculated using the PBE exchange correlation functional and PAW pseudopotentials. Self-consistent calculations of the electron density were performed on an $8 \times 8 \times 8$ Monkhorst-Pack $k$ grid with plane-wave energy cutoffs of 400 eV. To simulate the effect of pressure on the system, we kept the shape of the crystal unchanged ($a = b = 0.636c$) but adjusted its lattice constant ranging from $a = 4.42$ Å to $a = 4.02$ Å. As shown in Figs. 6(c)-6(e), the band inversion transition begins to occur when the lattice constant is around 4.22 Å and the system transforms from the NP to the TP. In virtue of Wannier90 [56] and Wannier tools packages [57], we can get the maximally localized Wannier function (MLWF), which allows us to conveniently construct the dipole moment elements [58] and shift vectors.

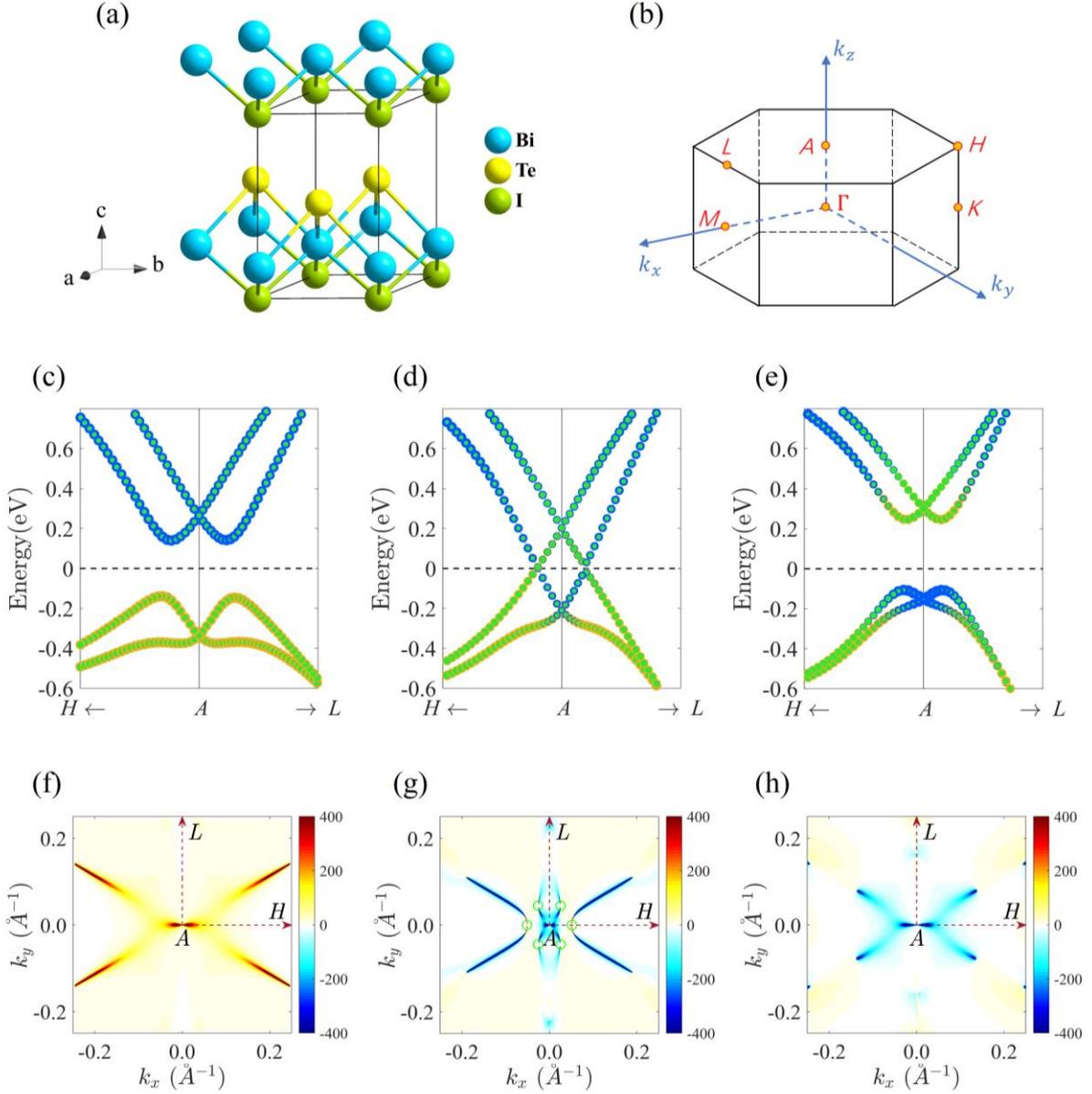

FIG. 6. (a) Layered crystal structure of BiTeI and its primitive cell (the black frame). (b) The corresponding Brillouin zone. The effects of changing the lattice constants that start from $a = 4.42$ Å (c,f), to $a = 4.22$ Å (d,g), to $a = 4.02$ Å (e,h). The bandgap undergoes the process of closing and reopening around the $A$ point. The orbital compositions reverse near the minimum bandgap (c-e), where the blue, yellow, green dots represent the weight from Bi-$6p$, Te-$5p$ and I-$5p$ atomic orbits in the Bloch wave functions, respectively. The black dashed line represents the Fermi energy. (f-h) As the band inversion occurs, the shift vector simultaneously reverses and which starts at the points where the bandgap closes. In (g), the six small green circles mark three pairs of Weyl points along the $A$-$H$ direction where the bandgap closing occurs.

Figures 6(f)-6(h) show shift vectors along the $k_x$ direction (the applied laser field is in the same direction) on the $LAH$ surface between the lowest conduction band and the highest valence band. The shift vector is invarient under the time reversal and mirror reflection about $\Gamma AL$ plane, and the mirror is parallel to the shift vector. Figures 6(c)-6(h) manifest the reversal of the shift vector that results from the band inversion. When the phase transition occurs, due to the singularity property of Weyl points, the shift vectors at the Weyl points are divergent [59].

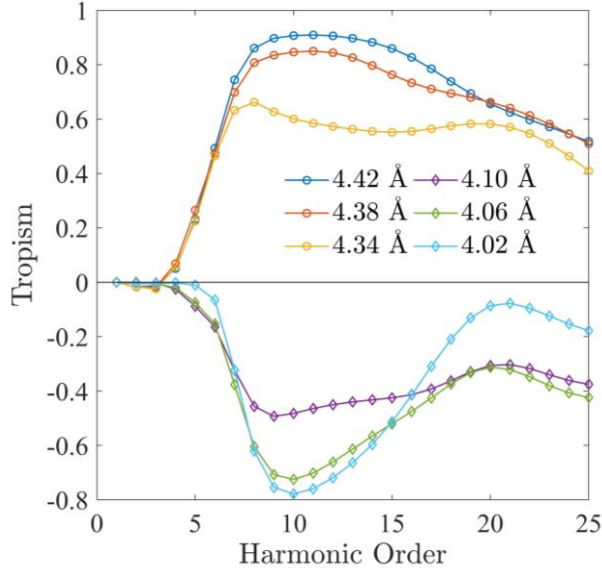

FIG. 7. HHG tropisms to the laser field in BiTeI with different lattice constants. Two lowest conduction bands and two highest valence bands with Rashba splitting are considered for calculations. Both the laser and nonlinear effect that we focus on are oriented along the $k_x$ direction. The diagrams of the NP and TP are drawn by lines with circles and diamonds, respectively. The two phases can be directly distinguished by signs of the tropisms. We do not plot the case approaching to the phase transition, because the band inversion region as well as the bandgap are too small to present significant tropisms.

Finally, we draw the tropisms of harmonics from BiTeI in a strong laser field. Since we only focus on low energy excitation near the Fermi energy, the driving laser of wavelength $\lambda = 15.6\ \mu m$ and peak strength $E_0 = 1.54\ MV/cm$ is used. For simplicity, we only consider the HHG contribution from the $LAH$ surface in the BZ. A $300 \times 300$ meshy $k$-point sample over the surface is chosen to simulate the time dependent evolution. From Fig. 7, we can see that there is an obvious sign reverse between tropisms of different phases. The spectral lines are positive in the NP, and negative in the TP. Since their cutoff frequencies are around the 25th order, we present the tropism of HHG from the 1st to the 25th order. The magnitude and position of the spectral peaks are related to their corresponding shift vectors and band structures. As the bandgap increases, the peak position moves to the higher order region and the peak value tends to the limit of value 1 or −1 for tropism as defined. Consequently, the strong field nonlinear optical effect brought by the band inversion is very clear. This signifies that it is promising to detect the process of band inversion and topological phase transition in non-centrosymmetric materials by means of the ultrafast all-optical strong-field method.

## V. CONCLUDING REMARKS

In conclusion, our work improves the electron tunneling picture in real space driven by a strong laser, and reveals the key role of the shift vector in solid HHG from non-centrosymmetric system, which has been overlooked so far. We find that HHPTs display novel asymmetric phenomenon in the time domain due to the inversion symmetry breaking, and the shift vector is responsible for the microscopic mechanism in essence. Benefited from the fact that the band inversion causes shift vector reversing, the HHPTs shows completely opposite radiation time, and we propose an all-optical method to characterize the band inversion during the topological phase transition in non-centrosymmetric TIs.

Fortunately, it is experimentally feasible that a nonlinear cross-correlation measurement using an ultrashort gating pulse to scan the HHG with a time delay is able to retrieval the HHPTs [16,53]. Our work establishes a close link between the strong-field nonlinear optics and topological condensed matter physics.

In addition to the phenomena mentioned above, due to the temporal asymmetry of HHG from systems with broken inversion symmetry, we consider the shift vector should also play a role in the even order harmonic generation [16,17,51,60] and broadening the harmonics cutoff [61]. Besides the TIs, it is expected to apply our conception to search for the Weyl semimetals without inversion centers and materials with large shift currents. Throughout this work, only the linearly polarized light is considered, while it is further expected to exploit new strong-field nonlinear phenomena under circularly polarized light combined with topological geometry [28,37,38,59]. The injection current from weak field generated by a sudden change in electron velocity is also instructive [59,62].

## ACKNOWLEDGMENTS

This work was supported by NSF of China (11974185, 11704187, 11834004, 11925408, 11921004, 12188101), the Natural Science Foundation of Jiangsu Province (Grant No. BK20170032), Fundamental Research Funds for the Central Universities (No. 30920021153), the Project funded by China Postdoctoral Science Foundation (Grant No. 2019M661841), the National Key Research and Development Program of China (2018YFA0305700), the Strategic Priority Research Program of CAS (XDB33000000), and the K. C. Wong Education Foundation (Grant No. GJTD-2018-01). C. Q. is very thankful to Yuting Qian, Guanglu Yuan, Tong Wu, and Xiangyu Tang for their help and discussions at the early exploration stage.

**APPENDIX A: NUMERICAL CALCULATION OF CURRENTS**

By solving the SBEs, i.e., Eq. (1), we can obtain the time-dependent population of electrons under the adiabatic Houston representation. Before that, a set of orthonormal and complete Bloch basis was obtained by the MLWF with the determinate Wannier gauge. This gives smooth Bloch states between adjacent $k$-points that can be used to construct the dipole matrixes. The Bloch basis is determined by the initial Hamiltonian of the system, and then the time-dependent Houston basis can be obtained by adiabatic evolution of the Bloch basis in reciprocal space.

The total current can be divided into off-diagonal and diagonal elements, named as the interband and intraband currents, respectively,

$$J_{inter}(t) = -\sum_{K \in BZ} \sum_{n \neq m} \rho_{nm}^{k(t)}(t) p_{nm}^{k(t)}, n \neq m, \quad \text{(A1)}$$

$$J_{intra}(t) = -\sum_{K \in BZ} \sum_{n} \rho_{nn}^{k(t)}(t) p_{nn}^{k(t)}, \quad \text{(A2)}$$

where the momentum operator can be given by $\hat{p}(k) = \partial_k \hat{H}(k)$. When we use Eq. (1) to calculate the time-dependent electron density, the Bloch basis is converted to the Houston representation, and then the momentum matrix element takes the form $p_{nm}^{k(t)} = \langle u_n^{k(t)} | \hat{p} | u_m^{k(t)} \rangle$, which can be calculated by

$$p_{nm}^{k(t)} = i\left(\varepsilon_n^{k(t)} - \varepsilon_m^{k(t)}\right) d_{nm}^{k(t)}, n \neq m, \tag{A3}$$

$$p_{nn}^{k(t)} = \nabla_k \varepsilon_n^{k(t)}. \tag{A4}$$

The high-harmonic spectrum in frequency domain is obtained by the Fourier transforms of the current. By doing wavelet transform for the current, we can get its time-frequency analysis sprectrogram.

**APPENDIX B: THE INTERBAND CURRENT IN TWO-BAND APPROXIMATION**

Our theory is mainly based on the saddle point equations, but their derivation contains several important approximations. Let us start discussions from the single-particle time-dependent Schrödinger equation (TDSE) under the Houston representation,

$$\dot{a}_n^{k(t)}(t) = -i\varepsilon_n^{k(t)} a_n^{k(t)}(t) - iE(t) \cdot \sum_m d_{nm}^{k(t)} a_m^{k(t)}(t), \tag{B1}$$

in which, $a_n^{k(t)}(t)$ is the time-dependent coefficient of the wave function of system.

$$\Psi(r,t) = \frac{1}{N} \sum_{k \in BZ} \sum_n a_n^{k(t)}(t) u_n^{k(t)}(r), \tag{B2}$$

where $N$ is the total number of unit cells. We introduce a dynamic phase factor acting on $a_n^{k(t)}(t)$ as

$$a_n^{k(t)}(t) e^{i\int_{-\infty}^t \varepsilon_n^{k(\tau)} d\tau} = b_n^{k(t)}(t). \tag{B3}$$

Then the TDSE becomes

$$\dot{b}_n^{k(t)}(t) = -iE(t) \cdot \sum_m d_{nm}^{k(t)} b_m^{k(t)}(t) e^{i\int_{-\infty}^t \varepsilon_n^{k(\tau)} - \varepsilon_m^{k(\tau)} d\tau}. \tag{B4}$$

We can obtain the time-dependent density of electrons via $\bar{\rho}_{nm}(t) = b_n^*(t) b_m(t)$, so under the two-band approximation, we have

$$\dot{\bar{\rho}}_{vc}^{k(t)}(t) = -iE(t) \left[ \left( d_{cc}^{k(t)} - d_{vv}^{k(t)} \right) \bar{\rho}_{vc}^{k(t)}(t) + d_{cv}^{k(t)} w^{k(t)}(t) e^{i\int_{-\infty}^t \varepsilon_g^{k(\tau)} d\tau} \right], \tag{B5}$$

in which, $w^{k(t)}(t) = \bar{\rho}_{vv}^{k(t)}(t) - \bar{\rho}_{cc}^{k(t)}(t)$ denotes the population difference of electrons between the valence band and conduction band. When the nonadiabatic tunneling excitation occurs in a semiconductor, we can use the Keldysh approximation that $w^{k(t)}(t) \approx 1$ [46]. Thus, the interband and intraband densities can be decoupled. By solving the differential Eq. (B5) and using Eqs. (A1) and (A3), we can obtain the interband current as

$$J_{inter}(t) = \sum_{K \in BZ} \int_{-\infty}^{t} dt' \, \varepsilon_g^{k(t)} \left| d_{vc}^{k(t)} \right| E(t') \left| d_{cv}^{k(t')} \right| e^{-iS(K,t,t')} + c.c.. \tag{B6}$$

Formally taking its Fourier transform, we can get Eq. (2).

The most important point here is whether the Keldysh approximation is strictly satisfied. When the topological phase transition occurs, the bandgap of the system tends to close. This can lead directly to a surge of charge carriers around the bandgap, and Keldysh approximation is no longer satisfied at this time. Therefore, our theoretical frame cannot be extended to Dirac systems or topological semimetals [18,59].

**APPENDIX C: THE INDISPENSABILITY OF BREAKING THE INVERSION SYMMETRY**

Here, we show the reason why we only discuss the systems with broken inversion symmetry. The shift vector is composed of equilibrium state wave functions and can be used to describe the intrinsic microscopic property of systems. Considering a Bloch state of electron under spatial inversion transformation $P$, it transforms as

$$P|u_{n,k}\rangle \rightarrow |u'_{n,k}\rangle = |u_{n,-k}\rangle. \tag{C1}$$

For a crystal with inversion symmetry, its Bloch state is invariable under $P$ but should be multiplied by a phase factor,

$$P|u_{n,k}\rangle \rightarrow |u_{n,-k}\rangle = |u_{n,k}\rangle e^{i\phi_n^k}. \tag{C2}$$

Then we can get the relationship between the dipole matrix elements at $k$ and $-k$ points in the BZ.

$$\begin{aligned}
d_{nm}^k &= i\langle u_{n,k}|\nabla_k|u_{m,k}\rangle \\
&= i\left\langle u_{n,-k}\left|e^{i\phi_n^k}\nabla_k\left(e^{-i\phi_m^k}|u_{m,-k}\rangle\right)\right.\right. \\
&= ie^{i(\phi_n^k-\phi_m^k)}(-i\nabla_k\phi_m^k)\langle u_{n,-k}|u_{m,-k}\rangle + ie^{i(\phi_n^k-\phi_m^k)}\langle u_{n,-k}|\nabla_k|u_{m,-k}\rangle \\
&= e^{i(\phi_n^k-\phi_m^k)}\nabla_k\phi_m^k \delta_{n,m} - e^{i(\phi_n^k-\phi_m^k)}i\langle u_{n,-k}|\nabla_{-k}|u_{m,-k}\rangle \\
&= \nabla_k\phi_m^k \delta_{n,m} - e^{i(\phi_n^k-\phi_m^k)}d_{nm}^{-k}. \tag{C3}
\end{aligned}$$

For the diagonal elements, it reads

$$d_{nn}^k = \nabla_k\phi_n^k - d_{nn}^{-k}, \tag{C4}$$

which is the adiabatic Berry connection. The transition dipole phase (TDP) takes the form,

$$\phi_{nm}^k = \pi + \phi_n^k - \phi_m^k + \phi_{nm}^{-k}. \tag{C5}$$

Therefore, the shift vector transforms as

$$R_{nm}^k = d_{nn}^k - d_{mm}^k - \nabla_k\phi_{nm}^k$$

$$= \nabla_k \phi_n^k - d_{nn}^{-k} - \nabla_k \phi_m^k + d_{mm}^{-k} - \nabla_k(\pi + \phi_n^k - \phi_m^k + \phi_{nm}^{-k})$$

$$= -d_{nn}^{-k} + d_{mm}^{-k} + \nabla_{-k}\phi_{nm}^{-k}$$

$$= -R_{nm}^{-k}. \tag{C6}$$

We can see that when the system has the inversion symmetry, the shift vector is an odd function in the BZ, which vanishes when we integrate over the entire BZ.

Now we discuss the effect of the inversion symmetry on HHG. Using Eqs. (B7), the interband current from arbitrary quasi-crystal canonical momentum $K$ point can be written as follows

$$J_{inter}^K(t) = \int_{-\infty}^{t} dt'\, \varepsilon_g^{k(t)} \left|d_{vc}^{k(t)}\right| E(t') \left|d_{cv}^{k(t')}\right| e^{-i\int_{t'}^{t}\varepsilon_g^{k(\tau)}d\tau} e^{i\int_{k(t')}^{k(t)} R_{cv}^{k(\tau)} dk(\tau)} + c.c., \tag{C7}$$

in which, $k(t) = K + A(t)$ is the time dependent crystal momentum under the laser field. Here, we do not consider the influence from the asymmetry of driving field, but assume a periodically oscillating laser field with constant amplitude, so we have

$$E\left(t + \frac{T}{2}\right) = -E(t), \tag{C8}$$

where $T$ is the period of laser, and we also have

$$-K + A\left(t + \frac{T}{2}\right) = -k(t), \tag{C9}$$

where the $-K$ is the spatial inversion momentum of $K$. Therefore, the interband current from momentum $-K$ at time $t + \frac{T}{2}$ can be written as

$$J_{inter}^{-K}\left(t + \frac{T}{2}\right)$$

$$= \int_{-\infty}^{t} dt'\, \varepsilon_g^{k'\left(t+\frac{T}{2}\right)} \left|d_{vc}^{k'\left(t+\frac{T}{2}\right)}\right| E\left(t + \frac{T}{2}\right) \left|d_{cv}^{k'\left(t'+\frac{T}{2}\right)}\right| e^{-i\int_{t'+\frac{T}{2}}^{t+\frac{T}{2}}\varepsilon_g^{k'(\tau)}d\tau} e^{i\int_{k'\left(t'+\frac{T}{2}\right)}^{k'\left(t+\frac{T}{2}\right)} R_{cv}^{k'(\tau)} dk'(\tau)} + c.c$$

$$= -\int_{-\infty}^{t} dt'\, \varepsilon_g^{-k(t)} \left|d_{vc}^{-k(t)}\right| E(t) \left|d_{cv}^{-k(t')}\right| e^{-i\int_{t'}^{t}\varepsilon_g^{-k(\tau)}d\tau} e^{i\int_{-k(t')}^{-k(t)} R_{cv}^{k(\tau)} dk(\tau)} + c.c.$$

$$= -\int_{-\infty}^{t} dt'\, \varepsilon_g^{k(t)} \left|d_{vc}^{k(t)}\right| E(t) \left|d_{cv}^{k(t')}\right| e^{-i\int_{t'}^{t}\varepsilon_g^{k(\tau)}d\tau} e^{i\int_{k(t')}^{k(t)} R_{cv}^{k(\tau)} dk(\tau)} + c.c.$$

$$= -J_{inter}^K(t), \tag{C10}$$

where $k'\left(t + \frac{T}{2}\right) = -K + A\left(t + \frac{T}{2}\right)$. In the third step, we use the spatial inversion invariance of the band structure, and the TDAs are also spatial inversion invariant, which can be obtained from Eq. (C3), the shift vector is an odd function in the BZ that is considered as well. We can see that if the system has the inversion symmetry, the interband current satisfies $J_{inter}^{-K}\left(t + \frac{T}{2}\right) = -J_{inter}^K(t)$, this condition ensures pure odd-order harmonic generation. After integrating the entire BZ, the intensity of the current satisfies

$$I_{inter}\left(t + \frac{T}{2}\right) = I_{inter}(t), \tag{C11}$$

which is contrary to the temporal asymmetry of HHG we talked about in the main text.

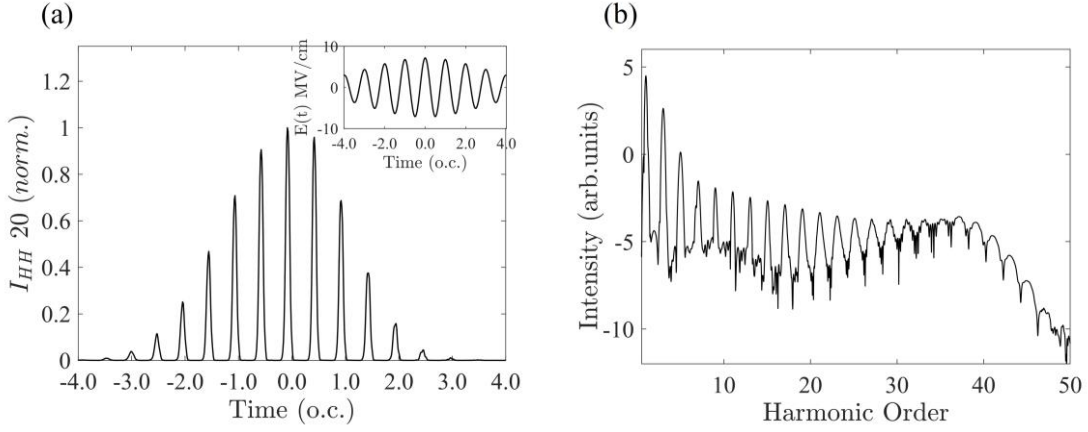

FIG. 8. HHG of the KM model with inversion symmetry. (a) The temporal structure of the 20th order HHPTs, the inset shows the corresponding laser field strength. (b) High-harmonic spectrum with pure odd-order harmonic generation.

Our numerical calculation results of the KM model with inversion symmetry breaking term $\lambda_v = 0$ are shown in Fig. (8). We can see that the HHPTs approximately satisfy Eq. (C11) but small deviation is due to the asymmetry of the driving field [Fig. 8(a)], this results in pure odd-order harmonic generation [Fig. 8(b)]. Our numerical results are in agreement with the theoretical derivation. Therefore, in order to get asymmetric signal in the time domain, the inversion symmetry breaking is indispensable in our discussions. Similarly, the mirror symmetry about the plane perpendicular to the laser field must also be broken, which can be achieved by changing the orientation angle of the laser field.

**APPENDIX D: THE TIME REVERSAL INVARIANCE OF THE SHIFT VECTOR**

As same as Eq. (C1), the Bloch state of electrons under time reversal $T$ transforms as

$$T|u_{n,k}\rangle \rightarrow |u'_{n,k}\rangle = \langle u_{n,-k}|. \tag{D1}$$

In this paper, the systems we considered possess the time reversal symmetry, so we have

$$T|u_{n,k}\rangle \rightarrow \langle u_{n,-k}| = |u_{n,k}\rangle e^{i\phi_n^k}. \tag{D2}$$

we can get the relationship between the dipole matrix elements at $k$ and $-k$ points in the BZ like the inversion symmetry case, as

$$d_{nm}^k = \nabla_k \phi_m^k \delta_{n,m} + e^{i(\phi_n^k - \phi_m^k)}(d_{nm}^{-k})^*. \tag{D3}$$

The Berry connection has

$$d_{nn}^k = \nabla_k \phi_n^k + d_{nn}^{-k}, \tag{D4}$$

and the TDP has

$$\phi_{nm}^k = \phi_n^k - \phi_m^k - \phi_{nm}^{-k}. \tag{D5}$$

At last, we can obtain the time reversal invariance of the shift vector,

$$\begin{aligned}
R_{nm}^k &= d_{nn}^k - d_{mm}^k - \nabla_k \phi_{nm}^k \\
&= \nabla_k \phi_n^k + d_{nn}^{-k} - \nabla_k \phi_m^k - d_{mm}^{-k} - \nabla_k (\phi_n^k - \phi_m^k - \phi_{nm}^{-k}) \\
&= d_{nn}^{-k} - d_{mm}^{-k} - \nabla_{-k} \phi_{nm}^{-k} \\
&= R_{nm}^{-k}. \tag{D6}
\end{aligned}$$

Although this symmetry simplifies our analyses, it is not essential to our conclusions, i.e., our research can be extended to magnetic materials without time reversal symmetry.

**APPENDIX E: THE CASE OF MULTIPLE LASER CYCLES**

Due to the inherent asymmetry of the laser field, the temporal asymmetry of the HHPTs could be modified by the laser conditions. The ideal laser field condition is described by Eq. (C8), here we consider the case of multiple laser cycles that nearly satisfies Eq. (C8). A driving field with the full-width at half-maximum of 240 fs is chosen with about 30 laser cycles involved, and other laser parameters are the same as those in Fig. 4(a). For the two phase points, we also show the HHPTs of the 20th and 25th orders in Fig. 9, other high-order cases follow the same pattern. Obviously, we get the same rule as Figs. 4(e) and 4(f), indicating that our conclusions are robust and do not come from the asymmetry of laser fields.

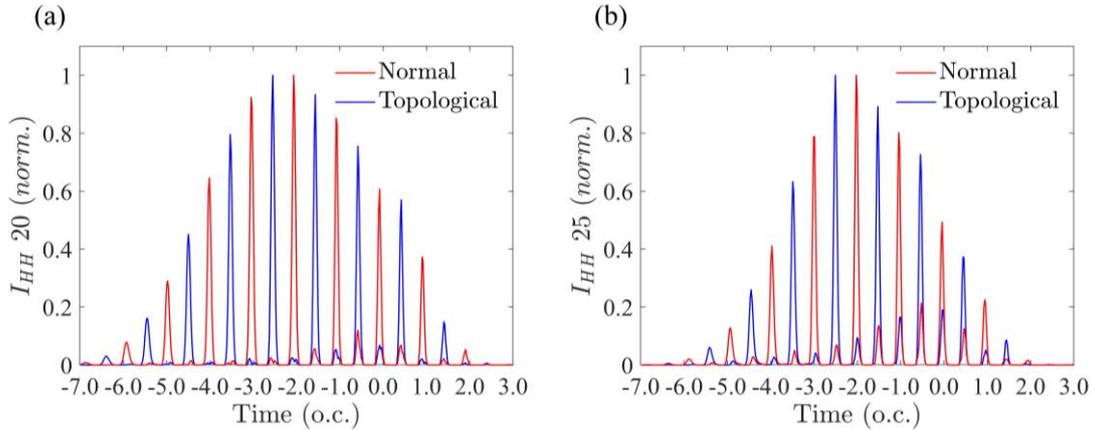

FIG. 9. Normalized intensities of (a) the 20th and (b) 25th order HHPTs for the NP and TP points under a 240-fs laser field. Due to large difference in radiation intensity, only 10 optical cycles with the obvious values are shown.

**APPENDIX F: DEPENDENCE OF HHG ON CARRIER-ENVELOPE PHASE IN TWO DIFFERENT PHASES**

For the frequency spectra shown in Fig. 4(b), there is no obvious difference between the two phase points, which is due to their similar band structures and TDAs. Here, we compare the carrier-envelope phase (CEP) dependence for the NP and TP points. Similar spectral structures but a CEP shift of π can be seen in Fig. 10, because their shift vectors are inverse (Fig. 3). The laser field can be reversed when its CEP changes by π, which is equivalent to the reversal of the shift vector for the two phase points. Therefore, we see the difference between the topological and trivial phases from HHG in frequency domain, which can be realized by experiments conveniently.

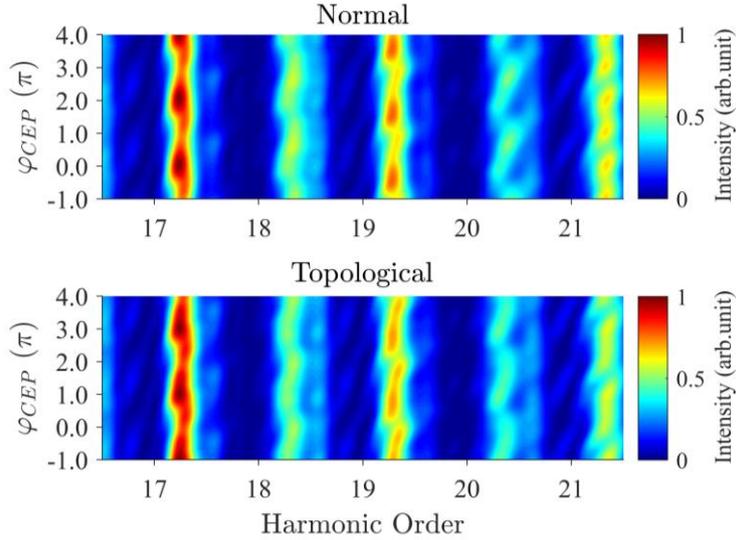

FIG. 10. CEP dependence of normalized HHG spectra for the NP and TP points. The CEP shift of π between the two spectral structures is originated from the reversal of the shift vector.

Moreover, it can be seen that the spectral lines have a significant deviation with respect to the integer orders, but there is no chirp in our driving lasers. Let us note from Fig. 4(c) that the high-order harmonics tend to radiate at the left half of laser fields, which is due to the special band structures of our model with small bandgaps and similar Dirac cones. There is a high probability of electron-hole pairs creating near the bandgap, and their early large accumulation reaches excitation saturation, which suppresses further formation of interband polarization. Therefore, the interband current will be mainly generated at the rising part of the laser field, where there is a positive amplitude chirp with a Gaussian envelope. Such increase in amplitude can be equivalent to a positive chirp in the laser field [63,64], thus the frequency of harmonics is no longer integer multiples of driving photons. Consequently, a blue shift of HHG can be observed in solids with small bandgaps as well as similar Dirac cones.

**APPENDIX G: COMPARISON OF INTERBAND AND INTRABAND RADIATIONS**

When we use saddle point approximation to do semi-classical analysis, only the interband contribution is considered, and we conclude that the shift vector can affect the radiated photon energy when the interband recollision occurs. To confirm the reasonability of our method, we compare contributions from interband and intraband in the case of the NP point. Figure 11 shows that the HHG intensity of low-order region is dominated by the intraband contribution, but high-order region is dominated by the interband process, which is in agreement with Refs. [17,46]. The total HHG has a clear cutoff area, which is the result of destructive interference between the interband and intraband parts. We consider this comes from band structures with similar Dirac cones and linear dispersions, Ref. [18] gives a similar calculation result from Dirac systems. The same conclusion can be obtained from the TP point.

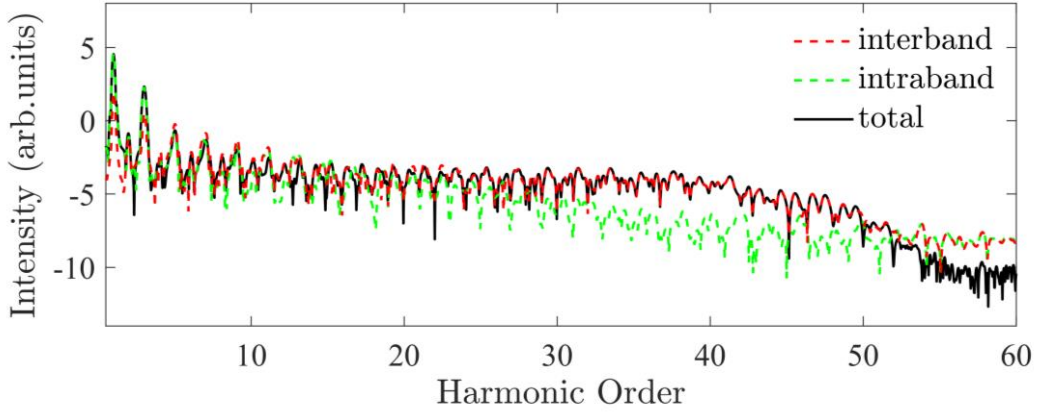

FIG. 11. High-harmonic spectra from the NP point in the KM model. The intraband and interband harmonics dominate the lower and higher frequency regions, respectively.

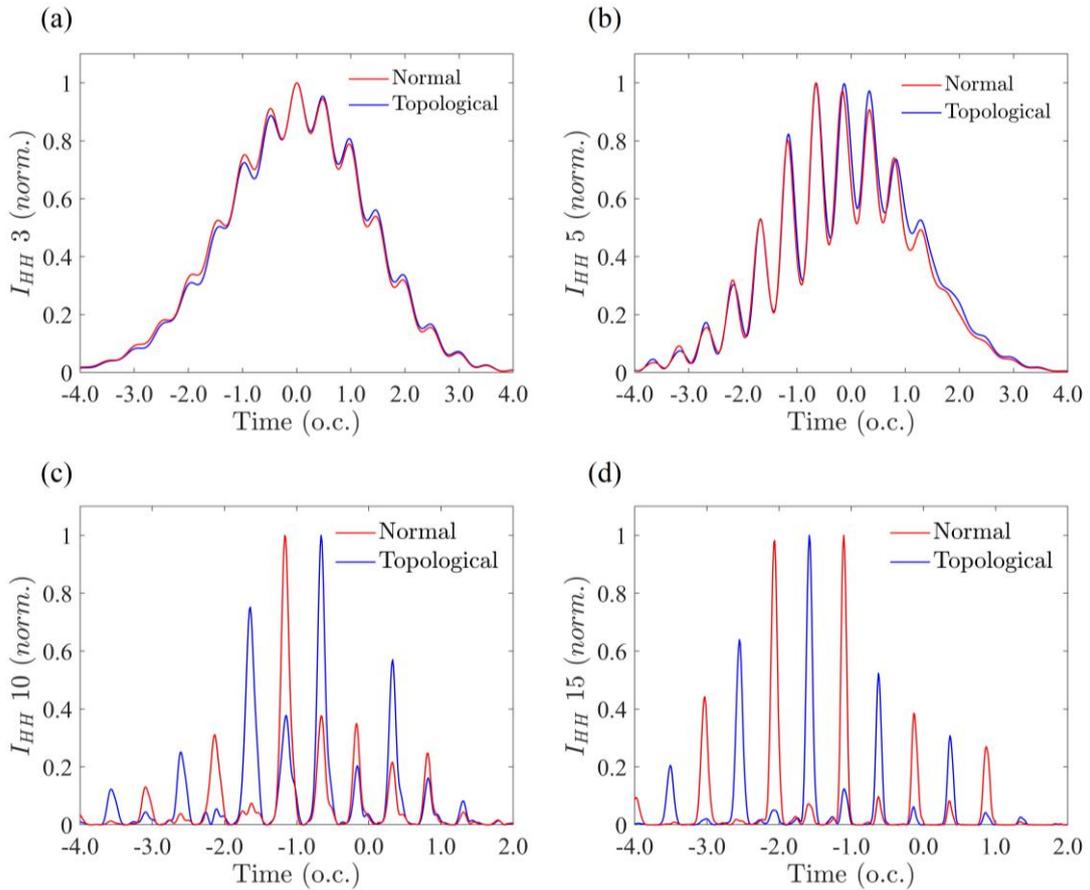

FIG. 12. HHPTs of the (a) 3th, (b) 5th, (c) 10th, and (d) 15th orders for the NP and TP phase points in the KM model. The laser condition is as same as Fig. 4.

In Fig. 12, we present several HHPTs of total current with lower orders relative to Figs. 4(e) and 4(f). The phenomenon of temporal asymmetry cannot be observed in the first few low-order harmonics, which are dominated by the intraband contribution. However, at higher harmonic orders (about $n \geq 15$), both the temporal asymmetry and the opposite radiation time between the NP and TP points are prominent. As the harmonics order increases, the contribution from interband catches up with that of intraband and gradually dominates the total harmonics[Fig. 11]. Therefore, we can see that these phenomena are indeed induced by the interband process, and our saddle point analysis only for the interband current is reasonable here. Of course, when strong coupling occurs between interband and intraband currents, such as behaviors in Dirac systems, the saddle point analysis is no longer practical.